%% file: paper.tex
\documentclass[twocolumn]{aastex63}
\usepackage{amsmath}
\usepackage{amssymb}
\usepackage{mathrsfs}

\definecolor{purple}{rgb}{0.5,0,0.5}
\definecolor{darkgreen}{rgb}{0.1,0.6,0.1}
\definecolor{orange}{rgb}{1,0.6,0}

\submitjournal{ApJ}
\shorttitle{Missing Convection Zones}
\shortauthors{Jermyn et al.}

\begin{document}

\title{A Transparent Window into Early-Type Stellar Variability}

\correspondingauthor{Adam S. Jermyn}
\email{adamjermyn@gmail.com}

\author[0000-0001-5048-9973]{Adam S. Jermyn}
\affiliation{Center for Computational Astrophysics, Flatiron Institute, New York, NY 10010, USA}

\author[0000-0002-3433-4733]{Evan H. Anders}
\affiliation{CIERA, Northwestern University, Evanston IL 60201, USA}

\author[0000-0001-5048-9973]{Matteo Cantiello}
\affiliation{Center for Computational Astrophysics, Flatiron Institute, New York, NY 10010, USA}
\affiliation{Department of Astrophysical Sciences, Princeton University, Princeton, NJ 08544, USA}

\begin{abstract}
Subsurface convection zones are ubiquitous in early-type stars.
Driven by narrow opacity peaks, these thin convective regions transport little heat but play an important role in setting the magnetic properties and surface variability of stars.
Here we demonstrate that these convection zones are \emph{not} present in as wide a range of stars as previously believed.
In particular, there are regions which 1D stellar evolution models report to be convectively unstable but which fall below the critical Rayleigh number for onset of convection.
For sub-solar metallicity this opens up a \emph{stability window}  in which there are no subsurface convection zones. For LMC metallicity this surface stability region extends roughly between $8M_\odot$ and $16M_\odot$, increasing to $8M_\odot$ -- $35M_\odot$ for SMC metallicity. Such windows are then an excellent target for probing the relative influence of subsurface convection and other sources of photometric variability in massive stars.
\end{abstract}


\input{intro}

\input{theory}

\input{results}

\input{conclusions}

\acknowledgments

The Flatiron Institute is supported by the Simons Foundation.
This research was supported in part by the National Science Foundation under Grant No. PHY-1748958.
We thank Dominic Bowman for suggesting the addition of Figure~\ref{fig:window_spec} and Daniel Lecoanet for advice on computing the critical Rayleigh number in rotating convection.


\software{
\texttt{MESA} \citep[][\url{http://mesa.sourceforge.net}]{Paxton2011,Paxton2013,Paxton2015,Paxton2018,Paxton2019},
\texttt{MESASDK} 20190830 \citep{mesasdk_macos,mesasdk_linux},
\texttt{matplotlib} \citep{hunter_2007_aa}, 
\texttt{NumPy} \citep{der_walt_2011_aa}}

\clearpage

\appendix

\input{mesa.tex}

\input{chandra.tex}

\bibliography{refs}
\bibliographystyle{aasjournal}

\end{document}

%% file: intro.tex
\section{Introduction}\label{sec:intro}

Subsurface convection zones are ubiquitous in early-type stars~\citep{Cantiello:2009,2019ApJ...883..106C}.
Driven by narrow opacity peaks~\citep{1993ApJ...408L..85S}, these thin convective regions transport little heat but play an important role in setting the magnetic fields~\citep{Cantiello:2011,2020ApJ...900..113J} and chemical mixing~\citep{1970ApJ...160..641M}.

Subsurface convection zones may also play a role in setting the photometric variability of early-type stars, though the precise connection between these is not well-understood.
In particular, one of the most exciting potential observational probes of these subsurface convection zones is stochastic low-frequency variability \citep[SLF,][]{Blomme:2011,2019A&A...621A.135B}, but there is active debate into the origin of the SLF phenomenon~\citep{2019ApJ...886L..15L}.
Subsurface convective motions form one possible explanation~\citep{2021ApJ...915..112C}, as do internal gravity waves excited by core convection~\citep{2019A&A...621A.135B} and wind mass loss~\citep{2021A&A...648A..79K}.

Here we demonstrate that these convection zones are \emph{not} present in as wide a range of stars as previously believed.
In particular, there are regions which 1D stellar evolution models report to be convectively unstable but which fall below the critical Rayleigh number for onset of convection.
This means that the radiative conductivity is large enough to turn off the convective instability in these regions.

We begin in Section~\ref{sec:theory} with an overview of the theory of stellar convection and the critical Rayleigh number.
In Section~\ref{sec:results} we compute stellar evolutionary tracks and use these to demonstrate that over certain mass and age windows some subsurface convection zones are subcritical.
Importantly, this leaves a metallicity-dependent \emph{stability window}  where there are no subsurface convection zones. This region opens at around $8 M_\odot$ and, for the metallicity of the Small Magellanic Cloud (SMC, $Z = 0.002$), extends up to $35 M_\odot$.
The stability window is then an excellent target for understanding how much of the SLF phenomenon is due to subsurface convection, as it should be possible to compare the SLF properties of stars inside and outside the window to see if there is any abrupt change, or if the properties better track the more gradual evolution in core properties.
We conclude with a discussion of this and other observational implications in Section~\ref{sec:discussion}.

%% file: theory.tex
\section{Theory}\label{sec:theory}

In 1D stellar evolution calculations, convection is assumed to be active whenever the Ledoux condition for instability holds, namely
\begin{align}
	\nabla_{\rm rad} > \nabla_{\rm ad} + \nabla_\mu \left.\frac{\partial \ln T}{\partial \ln \mu}\right|_{P,\rho}.
	\label{eq:s}
\end{align}
Here $\mu$ is the mean molecular weight,
\begin{align}
\nabla_\mu \equiv \frac{d\ln \mu}{d\ln P}
\end{align}
is the composition gradient,
\begin{align}
	\nabla_{\rm ad} \equiv \left.\frac{\partial \ln T}{\partial \ln P}\right|_s
\end{align}
is the adiabatic temperature gradient,
\begin{align}
	\nabla_{\rm rad} \equiv \frac{3\kappa L P}{64 \pi G M \sigma T^4}
\end{align}
is the radiative temperature gradient, $\kappa$ is the opacity, $\rho$ is the density, $L$ is the luminosity, $P$ is the pressure, $T$ is the temperature, $G$ is the gravitational constant, $M$ is the mass below the point of interest, and $\sigma$ is the Stefan-Boltzmann constant.

However there is an additional requirement for convection, namely that the Rayleigh number
\begin{align}\label{eq:Ra}
	\mathrm{Ra} \equiv \frac{g (\nabla_{\rm rad}-\nabla_{\rm ad}) \delta r^3}{\nu \alpha}\left(\frac{\delta r}{h}\right)\left(\frac{4-3\beta}{\beta}\right)
\end{align}
exceed the critical value $\mathrm{Ra}_{\rm crit}$~\citep[e.g.][]{1961hhs..book.....C,Shore:1992}.
The Rayleigh number expresses the importance of buoyancy compared to diffusive processes; when $\rm{Ra} < \rm{Ra}_{\rm{crit}}$, buoyant forces are not strong enough to overcome diffusive processes, and there is no convection.
Here 
\begin{align}
	\alpha \equiv \frac{16 \sigma T^3}{3\kappa c_p \rho^2}
\end{align}
is the thermal diffusivity, $c_p$ is the specific heat at constant pressure, $g$ is the acceleration of gravity, $\delta r$ is the thickness of the Ledoux-unstable layer, $\nu$ is the kinematic viscosity, $h \equiv -dr/d\ln P$ is the pressure scale height, and $\beta=P_{\rm gas}/P$ is the radiation parameter.
The factor depending on $\beta$ arises from the thermal expansion coefficient, and is equal to the ratio of the density susceptibility to the temperature susceptibility ($\chi_\rho / \chi_T$).
See Appendix~\ref{appen:visc} for details on how we compute $\nu$.

The critical Rayleigh number depends on the precise setup of the convection zone, including geometry, the shape of the opacity profile, rotation, and the diffusivities~\citep{2008JFM...601..317N,2018PhRvF...3b4801G,doi:10.1080/03091929.2014.987670}.
For simplicity, we adopt Chandrasekhar's theory~\citep[][Chapter 29]{1961hhs..book.....C}, and assume that convection occurs by either direct onset or overstability, whichever has a lower $\mathrm{Ra}_{\rm crit}$.
For non-rotating systems this produces $\mathrm{Ra}_{\rm crit} \sim 10^3$, while for rotating systems it is somewhat greater because rotation stabilizes against convection.
See Appendix~\ref{appen:chandra} for details of how we compute $\mathrm{Ra}_{\rm crit}$.

%% file: results.tex
\section{Results}\label{sec:results}

\begin{figure*}
\centering
\begin{minipage}{0.49\textwidth}
\includegraphics[width=\textwidth]{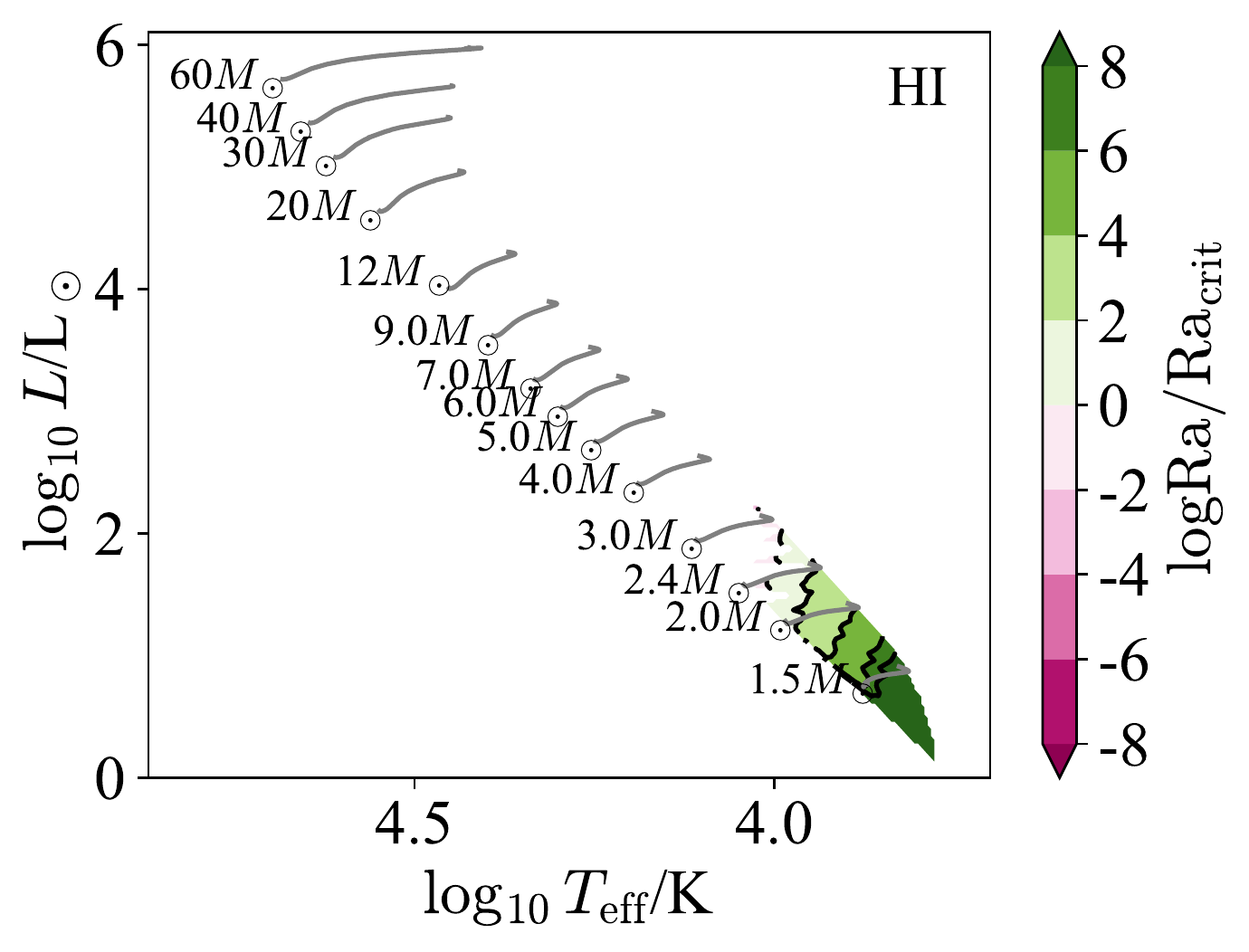}
\end{minipage}
\hfill
\begin{minipage}{0.49\textwidth}
\includegraphics[width=\textwidth]{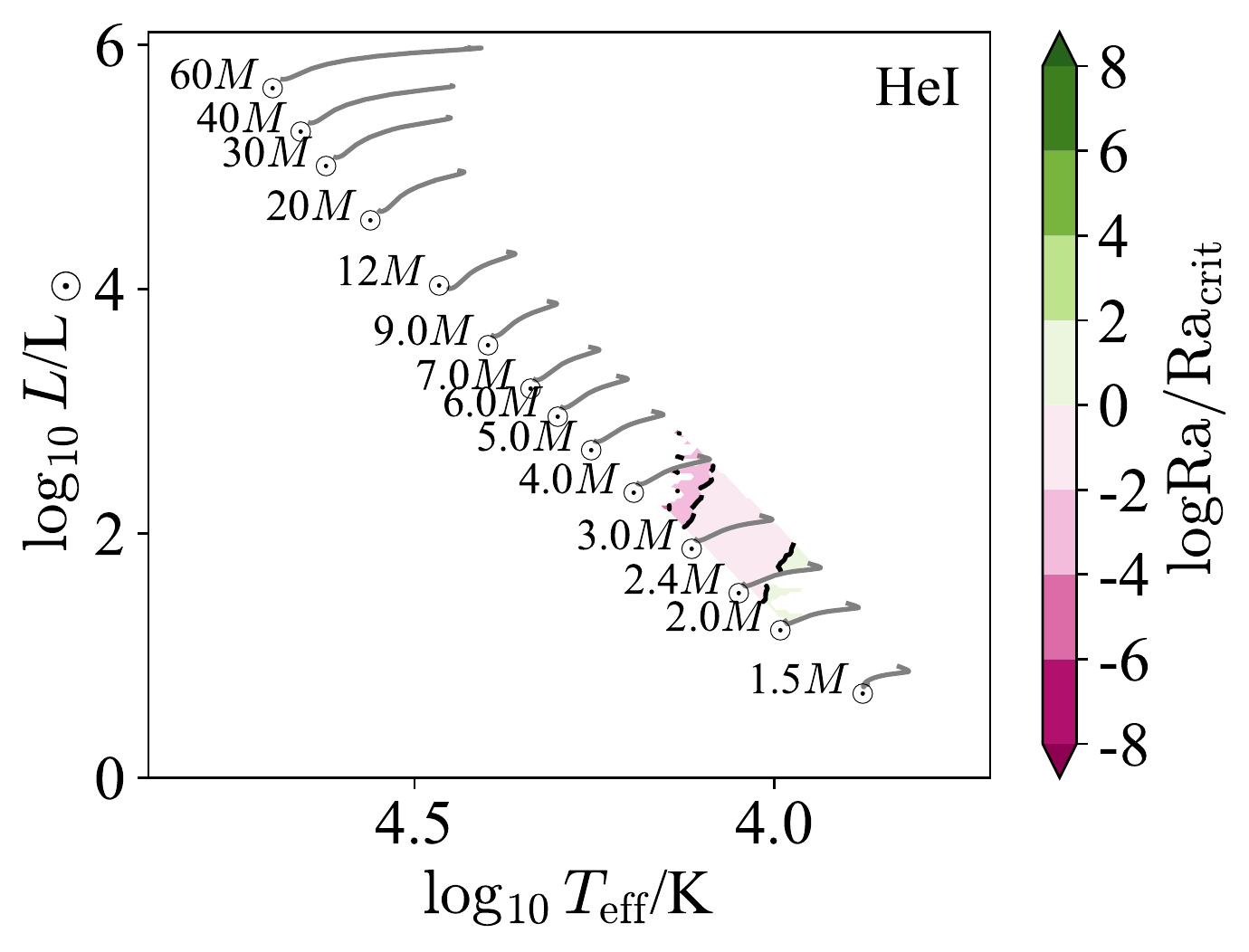}
\end{minipage}\\
\begin{minipage}{0.49\textwidth}
\includegraphics[width=\textwidth]{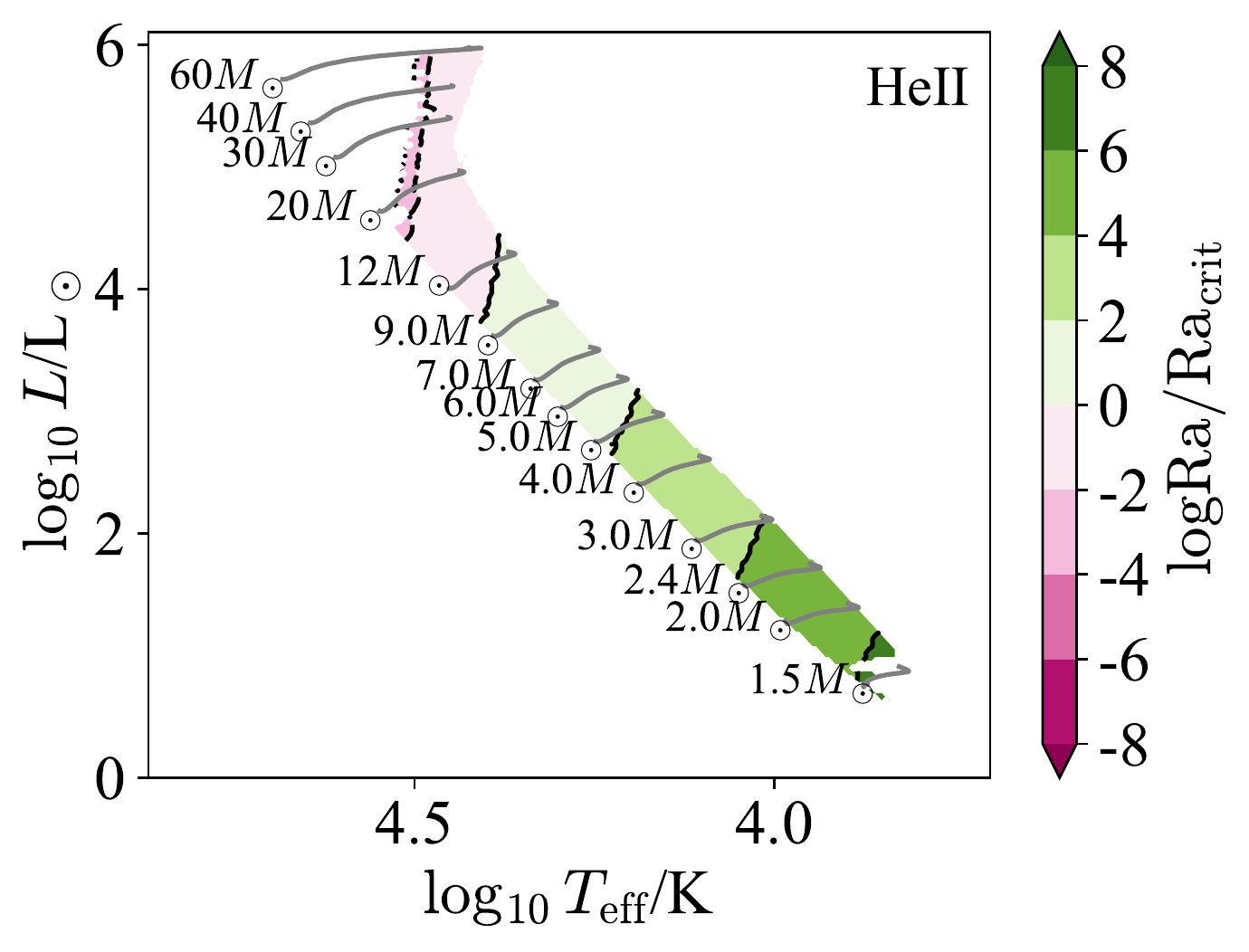}
\end{minipage}
\hfill
\begin{minipage}{0.49\textwidth}
\includegraphics[width=\textwidth]{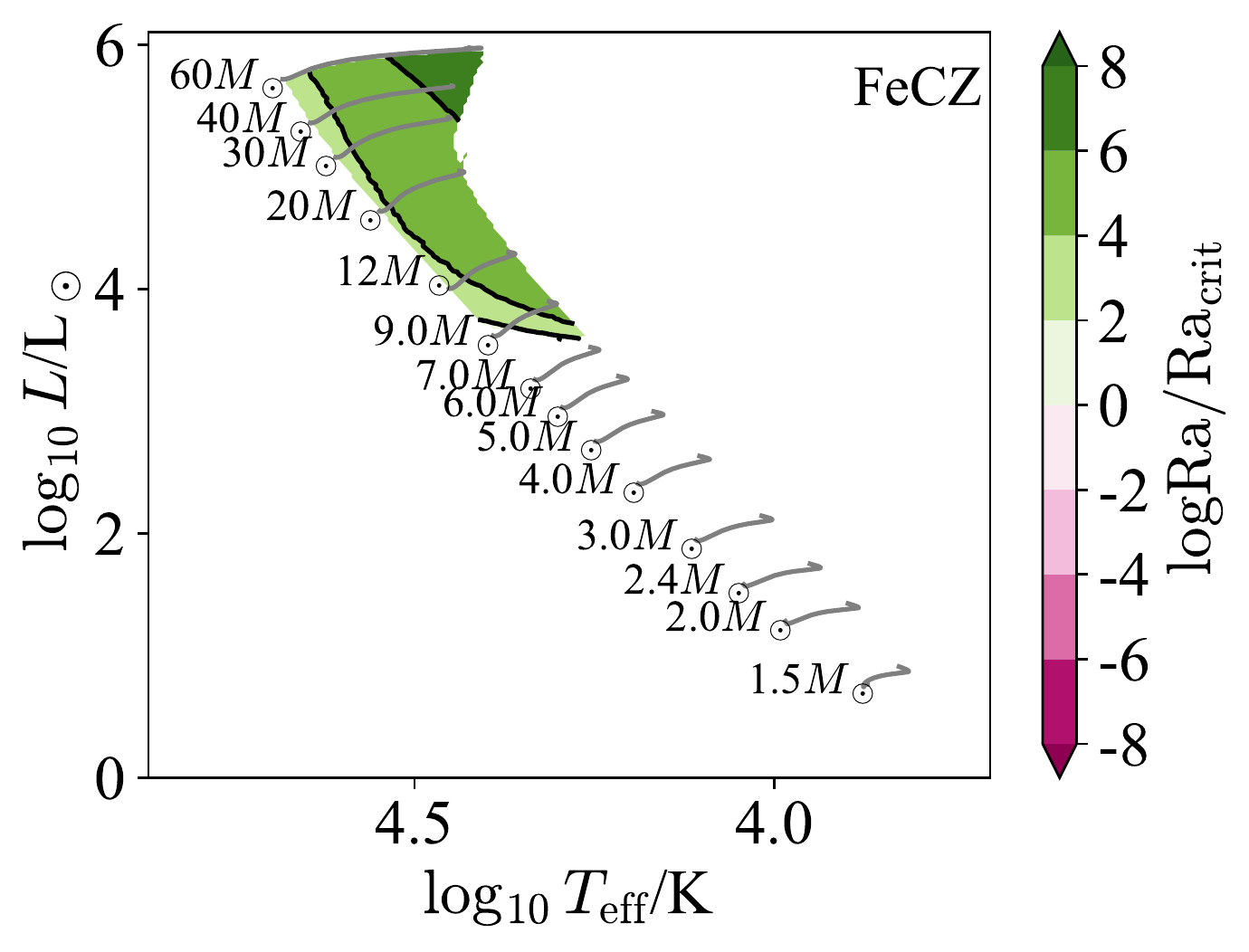}
\end{minipage}

\caption{The Rayleigh number normalized to the critical $\mathrm{Ra}_{\rm crit}$ is shown on a Hertzsprung-Russel diagram in terms of $\log T_{\rm eff}$ and $\log L$ for stellar models ranging from $1.1-60 M_\odot$ with Milky Way metallicity $Z=0.014$. The four panels correspond to four different subsurface convection zones, namely HI, HeI, HeII, and the FeCZ. White space indicates regions which lack a given convection zone in our models (i.e. equation~\eqref{eq:s} is not satisfied). To calculate $\mathrm{Ra}_{\rm crit}$ we assumed a rotation period of $10^3\,\mathrm{d}$.}
\label{fig:Ra}
\end{figure*}

We calculated stellar evolutionary tracks for stars ranging from $1.1-60 M_\odot$ using revision 15140 of the Modules for Experiments in Stellar Astrophysics
\citep[MESA][]{Paxton2011, Paxton2013, Paxton2015, Paxton2018, Paxton2019} software instrument.
Details on the MESA microphysics inputs are provided in Appendix~\ref{appen:mesa}.
Our models use convective premixing~\citep{Paxton2019} and determine the convective boundary using the Ledoux criterion.

We used equation~\ref{eq:Ra} to calculate the Rayleigh numbers in  subsurface convective layers caused by hydrogen, helium, and iron recombination (HI, HeI, HeII, and FeCZ). As they trace different ionization stages of H, He and Fe, these convective regions occur in specific temperature ranges \citep{2019ApJ...883..106C}. Moreover the FeCZ is only found in stars above a certain luminosity threshold, which in turns depends on the stellar metallicity \citep{Cantiello:2009}.

Figure~\ref{fig:Ra} shows the Rayleigh number normalized to the critical $\mathrm{Ra}_{\rm crit}$ ($\sim 10^3$, see Appendix~\ref{appen:chandra}) on a Hertzsprung-Russel diagram (HRD) for the four different subsurface convection zones with Milky Way metallicity $Z=0.014$.
A few features are striking.
First, the HeI convection zone is \emph{almost always} subcritical.
This is a missing convection zone: it is present in stellar models, and is still a superadiabatic layer in stars, but it likely does not undergo convective motions in most stars and has $\nabla = \nabla_{\rm rad}$ because convection transports no heat there.

Similarly the HI zone is subcritical for masses $2-3M_\odot$ and the HeII zone is subcritical above $5 M_\odot$.
The FeCZ by contrast is always supercritical where it exists.

\begin{figure*}
\centering
\includegraphics[width=\textwidth]{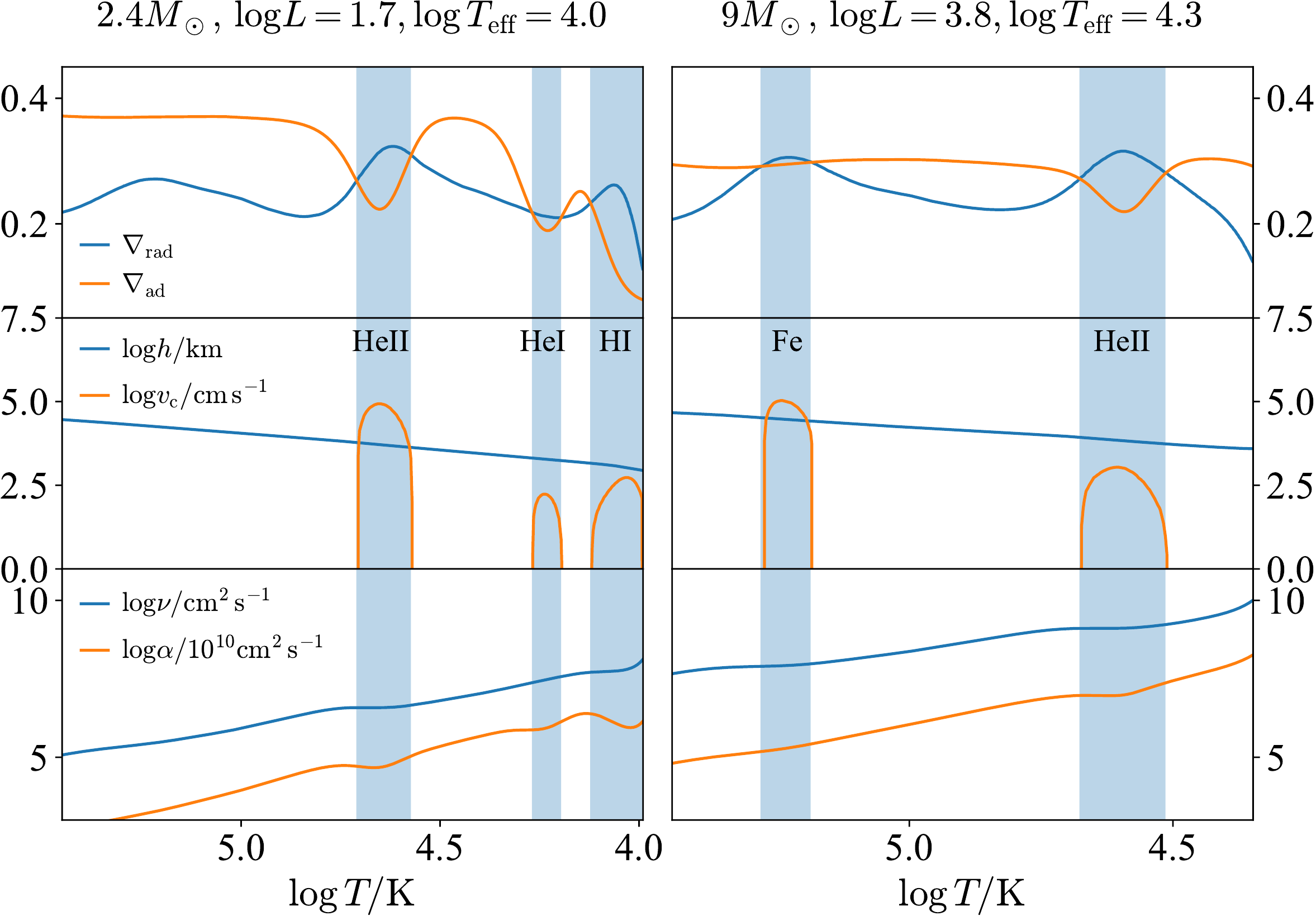}
\caption{Profiles of various quantities are shown for mid-main-sequence models of $2.4 M_\odot$ (left) and $9 M_\odot$ (right) stars at the Milky Way metallicity ($Z=0.014$) as functions of $\log T/\mathrm{K}$. The upper panels show the radiative ($\nabla_{\rm rad}$) and adiabatic ($\nabla_{\rm ad}$) temperature gradients. The middle panels show the pressure scale height $h \equiv P/\rho g$ and the convection speed computed by Mixing Length Theory. The lower panels show the viscosity $\nu$ and the thermal diffusivity $\alpha$.}
\label{fig:prof}
\end{figure*}

The differences in Ra between these different layers can be explained by a few factors.
First, the four CZs exist at very different densities.
At lower density photons have a longer mean free path, and since photons carry both heat and momentum this results in a higher thermal and kinetic diffusivity.
Examples of these trends are shown in the lower two panels of Figure~\ref{fig:prof}.
Moreover, the size of the convective region $\delta r$  is comparable to the local scale height, and this gets smaller for regions closer to the surface~\citep[][see also the middle panels of Figure~\ref{fig:prof}]{2019ApJ...883..106C}.

A careful examination of Figure~\ref{fig:Ra} reveals that there is \emph{always} a subsurface convection zone.
At Milky Way metallicity ($Z=0.014$) at least one of the four CZs is active for every mass we have examined and across the HRD.
This is not the case for all $Z$, however.

\begin{figure*}
\centering
\begin{minipage}{0.49\textwidth}
\includegraphics[width=\textwidth]{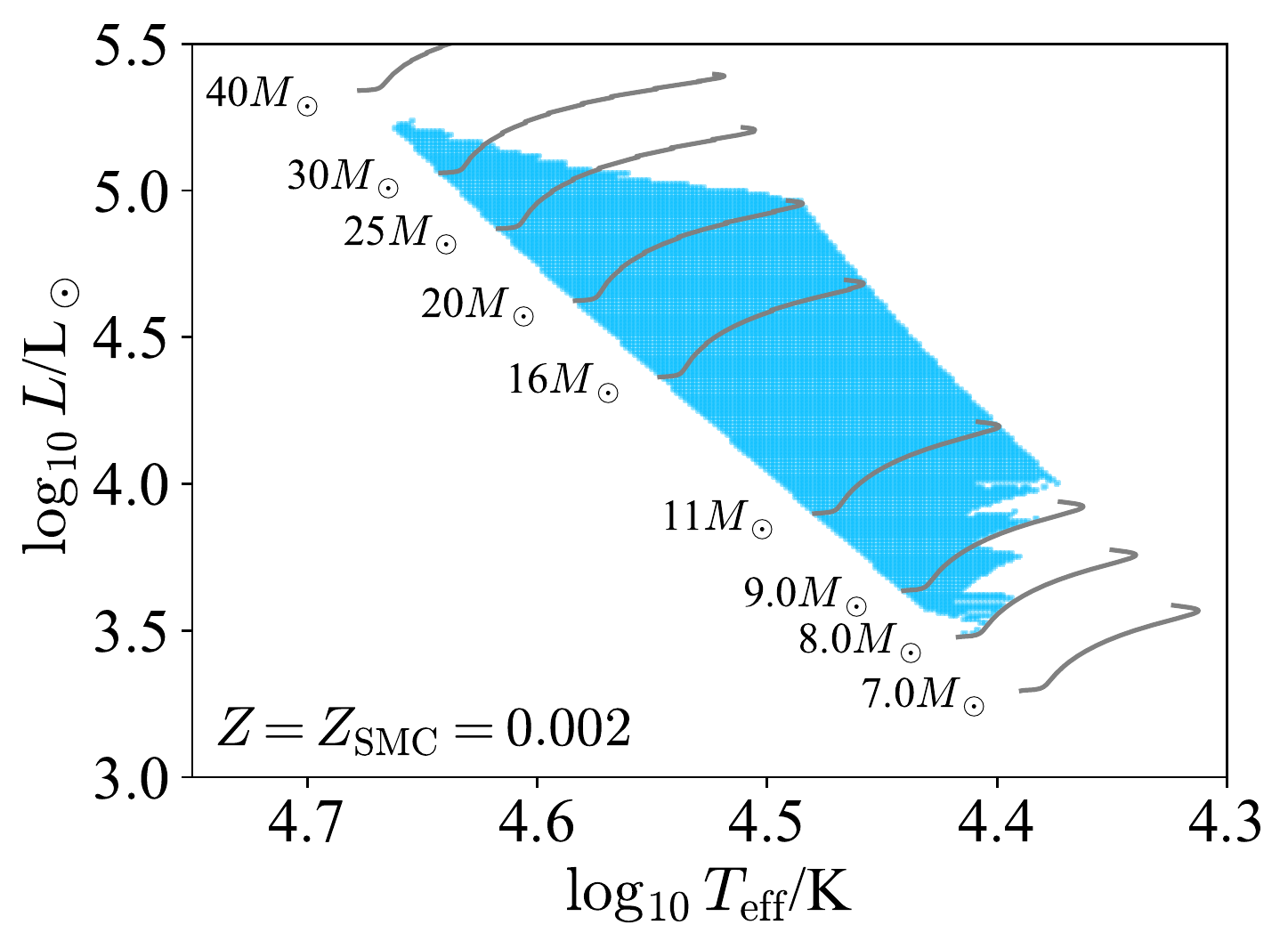}
\end{minipage}
\hfill
\begin{minipage}{0.49\textwidth}
\includegraphics[width=\textwidth]{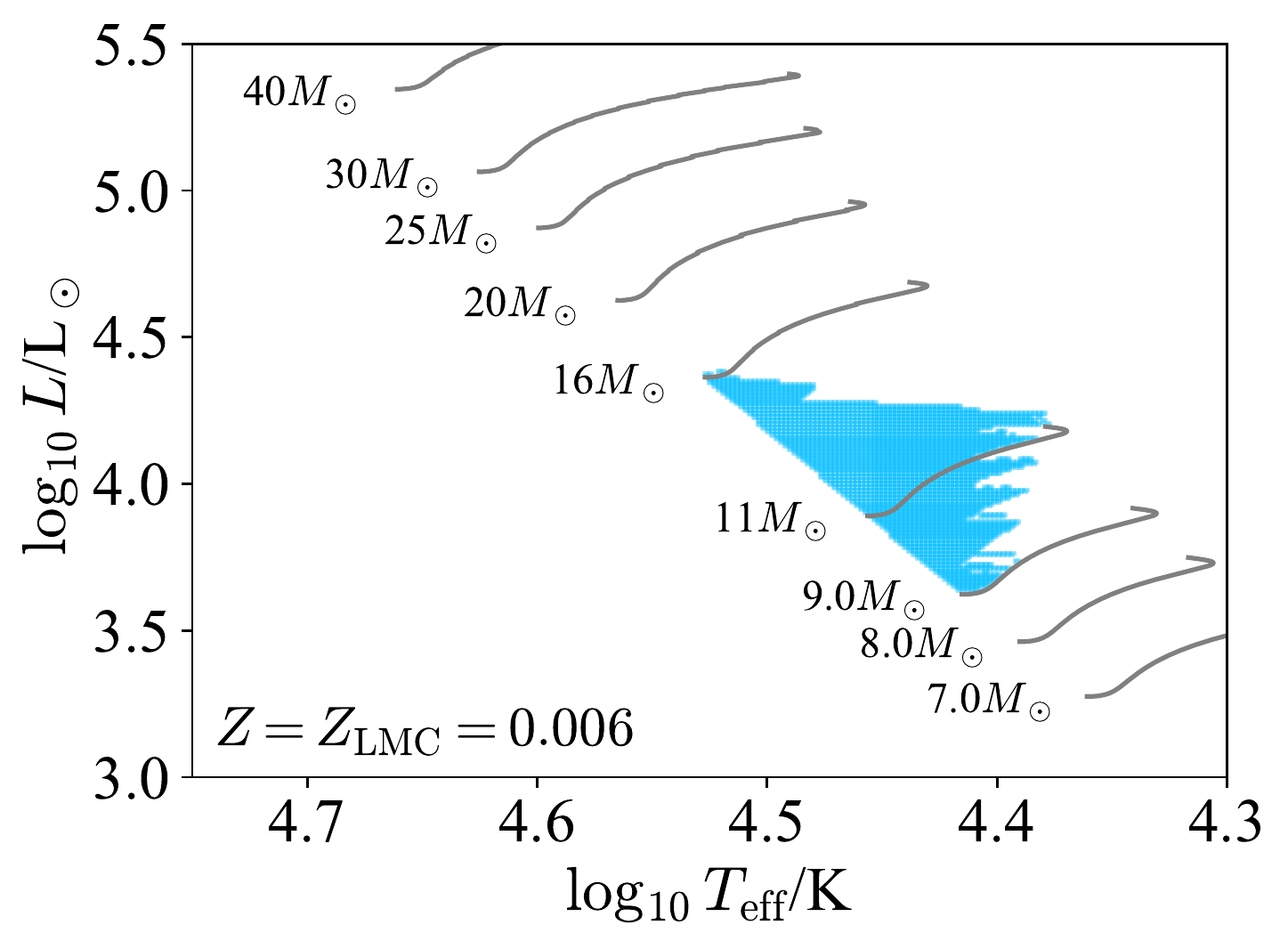}
\end{minipage}\\
\begin{minipage}{0.49\textwidth}
\includegraphics[width=\textwidth]{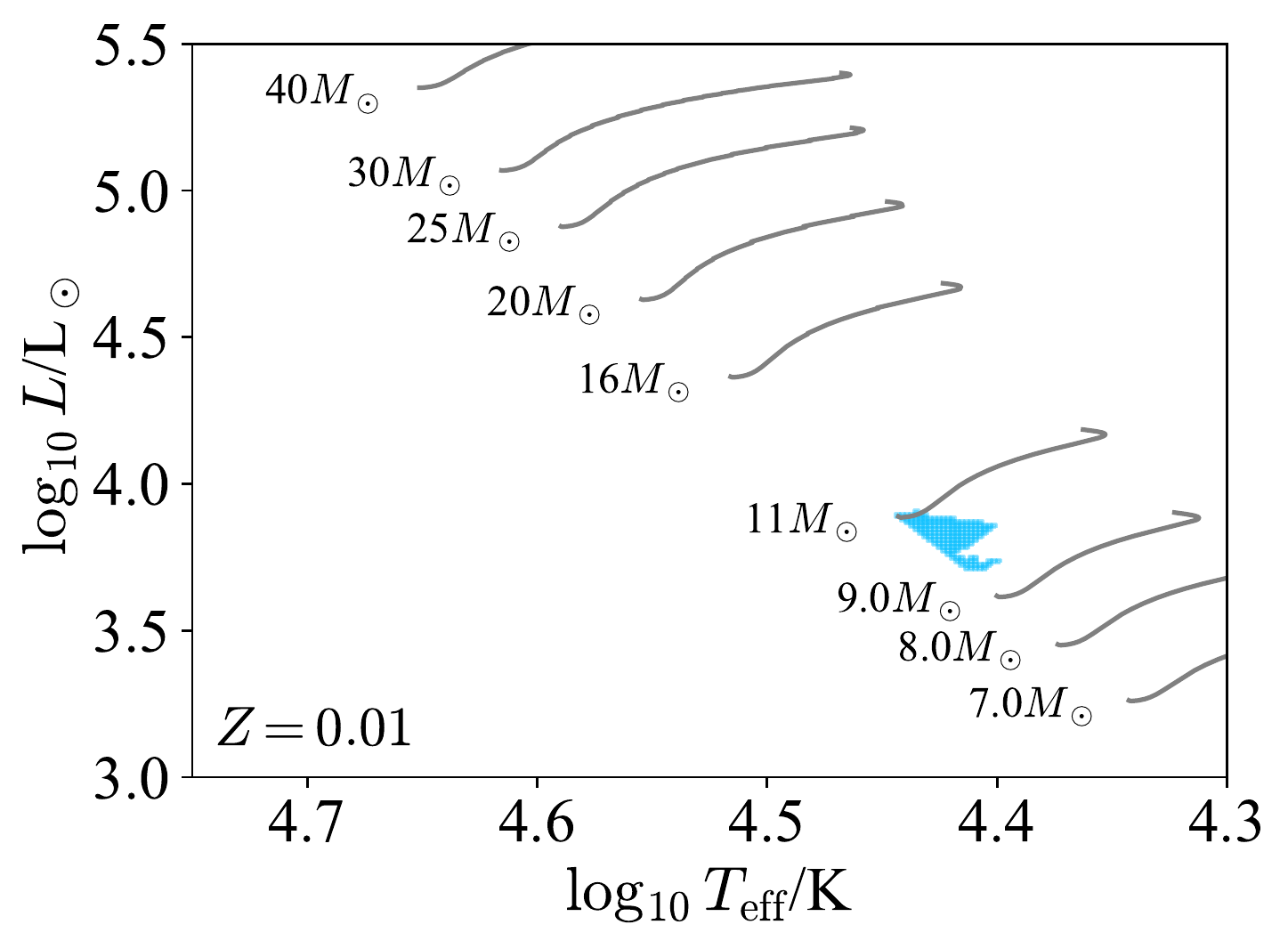}
\end{minipage}
\hfill
\begin{minipage}{0.49\textwidth}
\includegraphics[width=\textwidth]{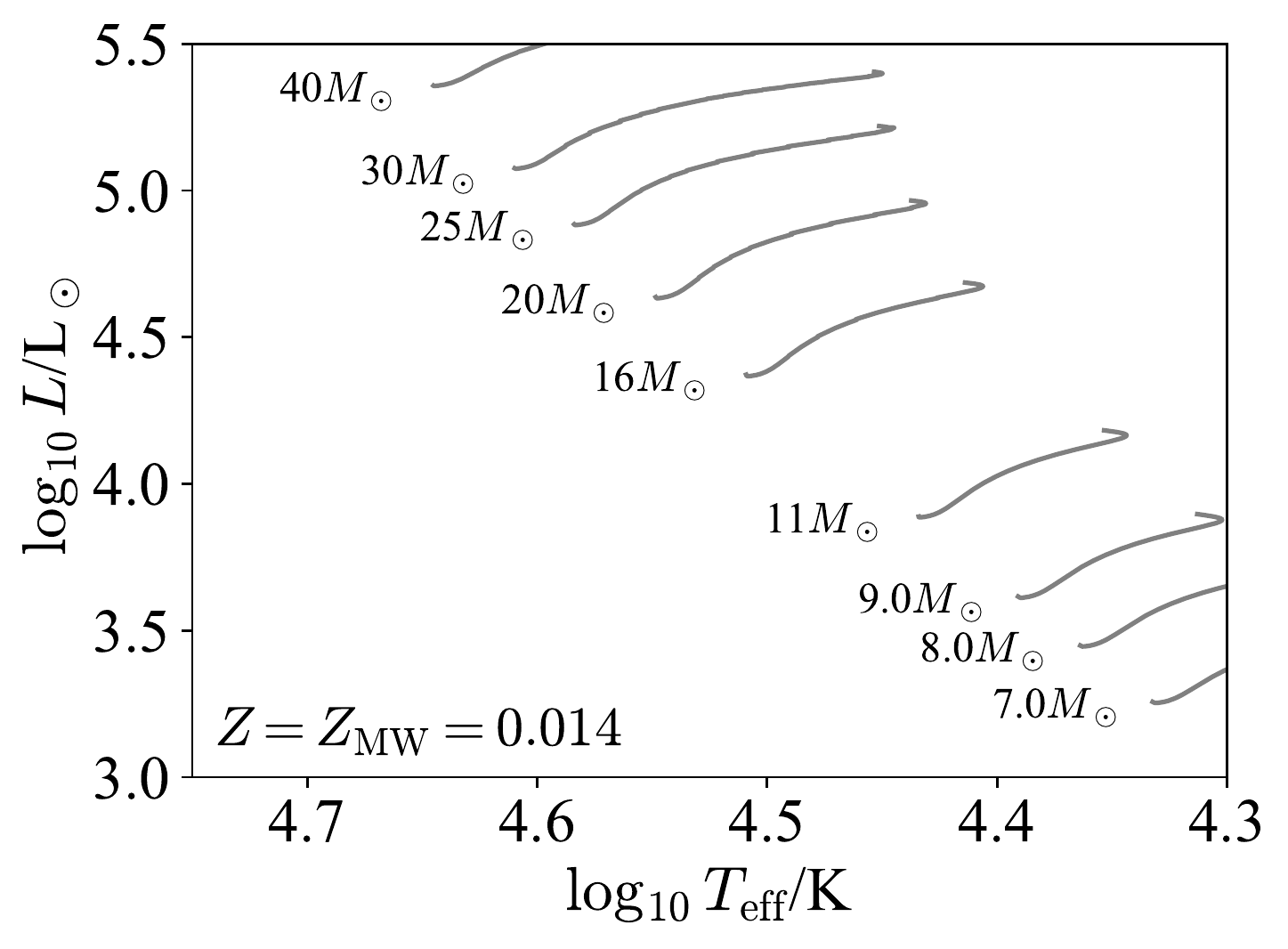}
\end{minipage}

\caption{The region where all subsurface superadiabatic layers have sub-critical Rayleigh numbers (the stability window) is shown on a Hertzsprung-Russel diagram in terms of $\log T_{\rm eff}$ and $\log L$ for stellar models ranging from $7-40 M_\odot$. The four panels correspond to four different metallicities, ranging from that of the Small Magellanic Cloud ($Z=0.002$) to the Milky Way ($Z=0.014$). At Milky Way metallicity and above the window with no subsurface convection disappears because the FeCZ moves down in $\log T_{\rm eff}$ and $L$ and meets the HeII CZ.  To calculate $\mathrm{Ra}_{\rm crit}$ we assumed a rotation period of $10^3\,\mathrm{d}$.}
\label{fig:window}
\end{figure*}

Towards smaller $Z$ the FeCZ moves up in $\log T_{\rm eff}$ and $\log L$ and a window opens where there are no subsurface convection zones.
This is shown in Figure~\ref{fig:window}, where we have plotted the window with no subsurface CZs for four different metallicities ranging from the Small Magellanic Cloud (SMC, $Z=0.002$) up to the Milky Way (where the window vanishes).
We call this the \emph{stability window}, because it is a region where subsurface convection is stabilized by microscopic diffusivities.

\begin{figure*}
\centering
\begin{minipage}{0.49\textwidth}
\includegraphics[width=\textwidth]{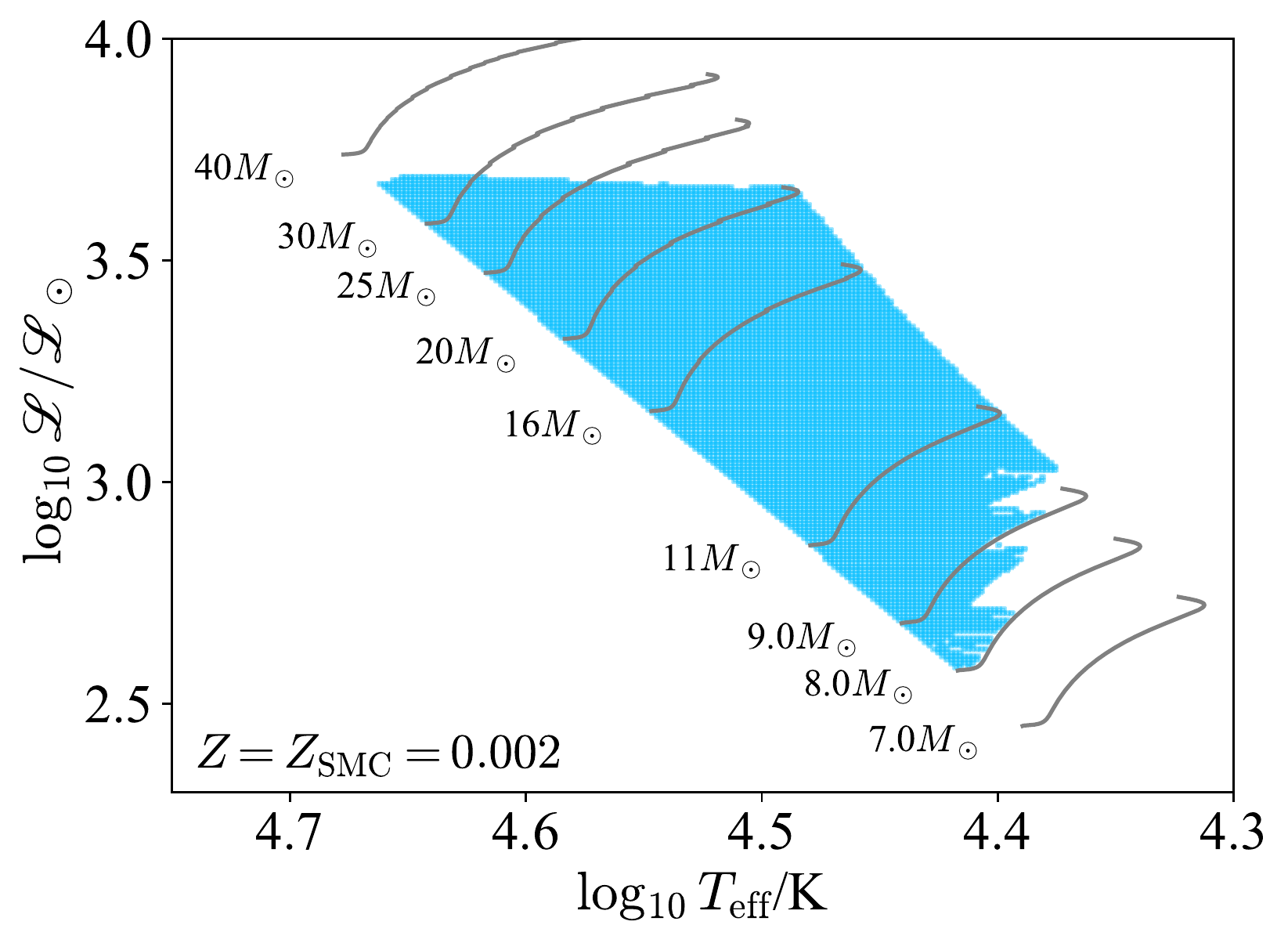}
\end{minipage}
\hfill
\begin{minipage}{0.49\textwidth}
\includegraphics[width=\textwidth]{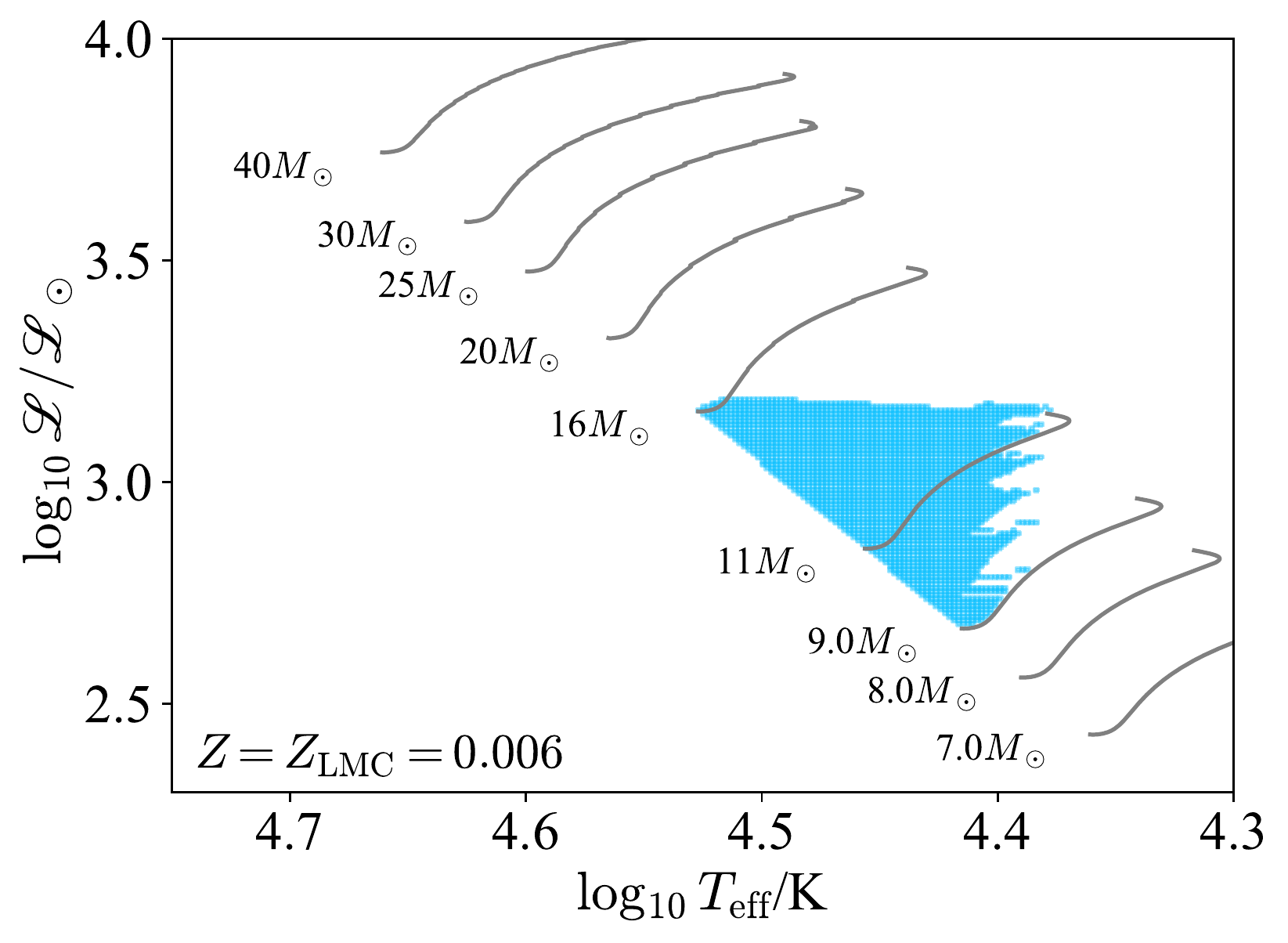}
\end{minipage}\\
\begin{minipage}{0.49\textwidth}
\includegraphics[width=\textwidth]{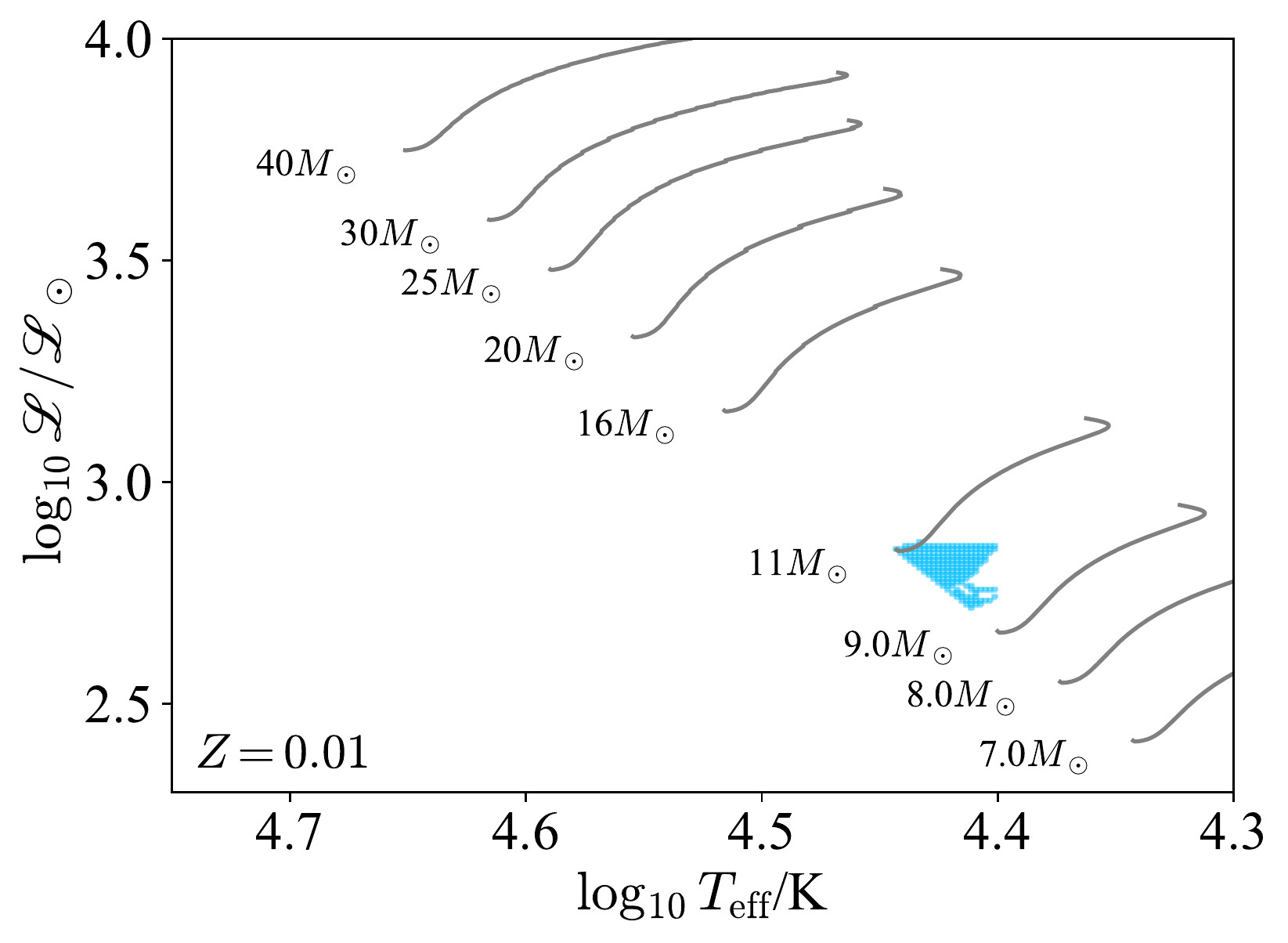}
\end{minipage}
\hfill
\begin{minipage}{0.49\textwidth}
\includegraphics[width=\textwidth]{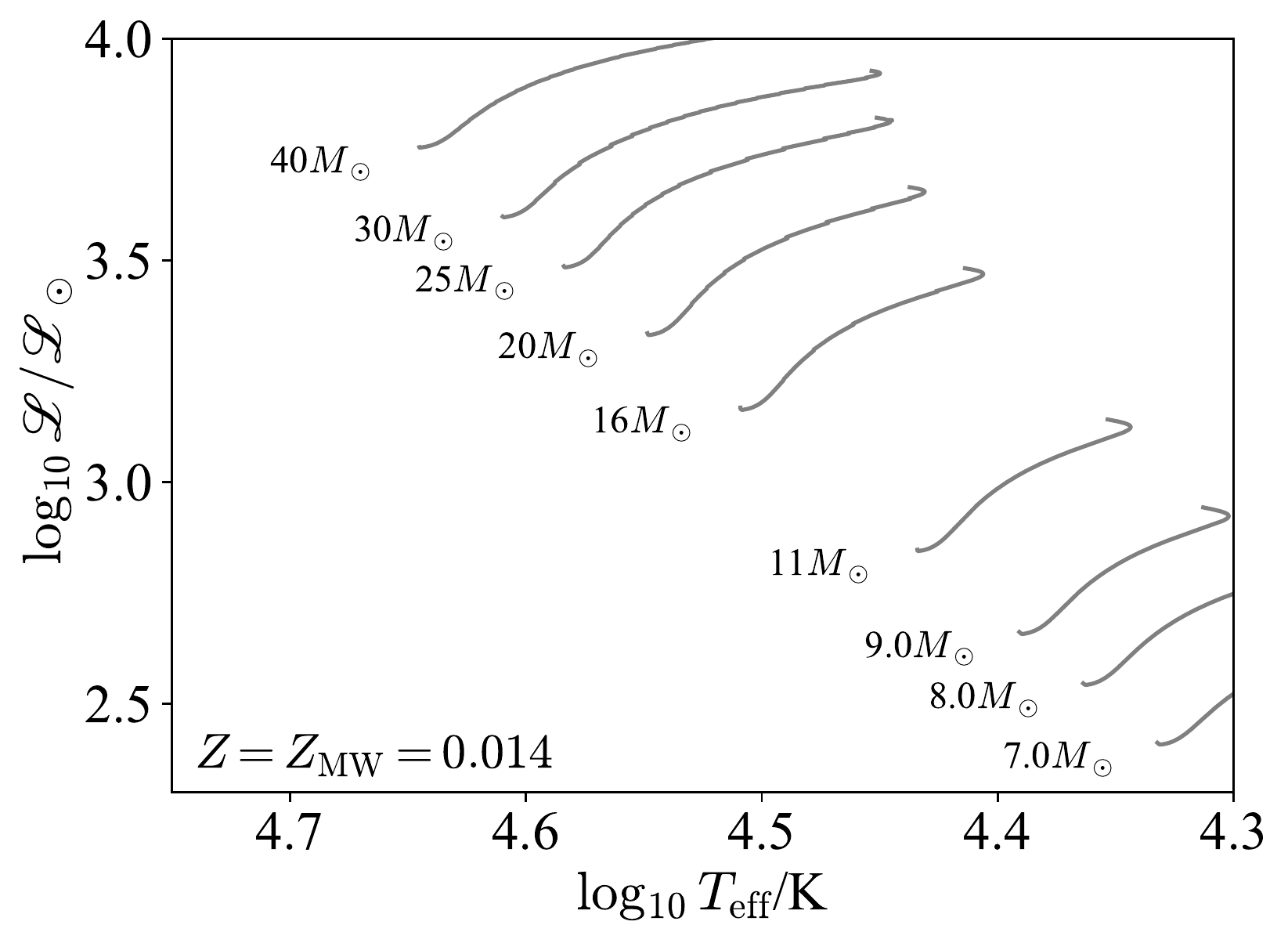}
\end{minipage}

\caption{The region where all subsurface superadiabatic layers have sub-critical Rayleigh numbers (the stability window) is shown on a Hertzsprung-Russel diagram in terms of $\log T_{\rm eff}$ and $\log \mathscr{L}$ for stellar models ranging from $7-40 M_\odot$. Here $\mathscr{L} \equiv L/M \propto T_{\rm eff}^4/g$ is the spectroscopic luminosity. The four panels correspond to four different metallicities, ranging from that of the Small Magellanic Cloud ($Z=0.002$) to the Milky Way ($Z=0.014$). At Milky Way metallicity and above the window with no subsurface convection disappears because the FeCZ moves down in $\log T_{\rm eff}$ and $L$ and meets the HeII CZ.  To calculate $\mathrm{Ra}_{\rm crit}$ we assumed a rotation period of $10^3\,\mathrm{d}$.}
\label{fig:window_spec}
\end{figure*}

At SMC metallicity the stability window is wide, from $8-35 M_\odot$ and covering the whole main-sequence for most of those masses.
As $Z$ rises the upper end of the window comes down, tracing the FeCZ, first to $16 M_\odot$ at Large Magellanic Cloud (LMC) metallicity ($Z=0.006$) then $10 M_\odot$ at $Z=0.01$.
Finally by Milky Way metallicity of $Z=0.014$ the start of the FeCZ pulls down to $8 M_\odot$, matching the upper mass at which HeII convection happens, and the window disappears.
For convenience the same quantities are shown in Figure~\ref{fig:window_spec} versus spectroscopic luminosity $\mathscr{L} \equiv T_{\rm eff}^4/g$, where $g$ is the surface gravity.

Rotation modifies our scenario in two ways.
First, it slightly changes the Rayleigh number in these subsurface superadiabatic zones by changing the hydrostatic structure of the star, but this is quite a small effect.
The reason for this is that these zones are generated by opacity bumps which occur at nearly fixed $T$.
Taking the density to be unchanged, and treating these zones as of order the pressure scale height $h$ in radial extent, we find
\begin{align}
	\mathrm{Ra}_{\rm rotating} \sim \frac{h^3 g_{\rm eff}}{\alpha \nu} \propto \frac{1}{g_{\rm eff}^2} \propto \frac{1}{(1-\Omega^2/\Omega_{\rm c}^2)^2},
\end{align}
where $g_{\rm eff}$ is the effective acceleration of gravity accounting for rotation, $\Omega$ is the angular velocity, and $\Omega_{\rm c}$ is the critical rotation rate.
Except very near critical rotation this is a small effect that does not compete with the many orders of magnitude variation in $\mathrm{Ra}$ across the HRD.

Secondly, and more importantly, rotation increases the critical Rayleigh number, stabilizing more stars against convection and growing the stability window.
In the calculations above we assumed a very slow rotation rate of $10^3\,\mathrm{d}$.
We do this because $\mathrm{Ra}_{\rm crit}$ has a different value when no rotation is assumed than when an extremely slow rotation is assumed.
Using slow but nonzero rotation rates slightly expands the stability windows.
The effects of increasing rotation further are small, however, and we see no difference in the stability windows calculated with rotation periods varying from $P=100\,\mathrm{d}$ down to $P=2\,\mathrm{d}$.

%% file: conclusions.tex
\section{Discussion}\label{sec:discussion}

How much do these missing convection zones matter?
For stellar structure, not much: subsurface convection zones carry very little heat, the temperature gradient $\nabla \approx \nabla_{\rm rad}$, so in the absence of convective heat transport the near-surface temperature profiles of these stars are likely not that different.
However, for more subtle measurements such as magnetism and photometric variability the properties of subsurface convection zones are critically important.

For instance in~\citet{2020ApJ...900..113J} and~\citet{2021arXiv211003695J} we made predictions of the interaction between surface magnetism and subsurface convection zones.
At Milky Way metallicities and above these are unaffected, but at lower metallicities (e.g. SMC and LMC) those predictions are significantly altered because there is a window in mass where no subsurface convection occurs (Figure~\ref{fig:window}), and we now expect:
\begin{enumerate}
\item There should be no magnetic desert for stars in the stability window because there is no subsurface convection that can erase a magnetic field.
\item A corollary is that stars should typically have stronger surface magnetic fields in the window than those just below the window in mass, because there can be no surface convective dynamo fields in these stars, only fossil fields, which tend to be stronger.
\item If Discrete Absorption Components \citep[DACs,][]{Kaper:1996} are generated by magnetic spots then DACs should be steady rather than time-varying in the stability window, because there is no subsurface convection present to generate a disordered magnetic field.
\item If wind clumping is seeded close to the stellar surface via convectively-driven fluctuations in density and velocity~\citep[e.g.][]{Puls:2006,Puls:2008,Cantiello:2009} then the clumping properties of stars in the window could be very different. 
\end{enumerate}

Similarly, the stability window provides an excellent opportunity to test theories of photometric variability in early-type stars.
In particular there is active debate over the origin of surface low-frequency variability~\citep{2019ApJ...886L..15L}.
We are aware of three proposed explanations: subsurface convective motions~\citep{2021ApJ...915..112C}, internal gravity waves excited by core convection~\citep{2019A&A...621A.135B} and wind mass loss~\citep{2021A&A...648A..79K}.
Stars inside the stability window, however, have no subsurface convection.
Any photometric variability observed in the window, then, cannot come from convection and must come from either one of the two other proposals or else some process not yet considered.
If the SLF phenomenon has multiple causes, we should expect a discontinuity on the edge of the window, and if it is monocausal we should either expect it to be indifferent to the window (ruling out a convective origin) or to be excluded from the window (requiring a convective origin).

%% file: mesa.tex
\section{MESA} \label{appen:mesa}

The MESA EOS is a blend of the OPAL \citep{Rogers2002}, SCVH
\citep{Saumon1995}, FreeEOS \citep{Irwin2004}, HELM \citep{Timmes2000},
and PC \citep{Potekhin2010} EOSes.

Radiative opacities are primarily from OPAL \citep{Iglesias1993,
Iglesias1996}, with low-temperature data from \citet{Ferguson2005}
and the high-temperature, Compton-scattering dominated regime by
\citet{Buchler1976}.  Electron conduction opacities are from
\citet{Cassisi2007}.

Nuclear reaction rates are from JINA REACLIB \citep{Cyburt2010} plus
additional tabulated weak reaction rates \citet{Fuller1985, Oda1994,
Langanke2000}.
Screening is included via the prescription of \citet{Chugunov2007}.
Thermal neutrino loss rates are from \citet{Itoh1996}.

Models were constructed on the pre-main sequence with $Y=0.24+2Z$ and $X=1-Y-Z$ and evolved from there.
We neglect rotation and associated chemical mixing.

\section{Viscosity}\label{appen:visc}

Computing the viscosity of a plasma is complicated.
For simplicity we use the viscosity of pure hydrogen plus radiation, so that
\begin{align}
	\nu \approx \nu_{\rm H} + \nu_{\rm rad}.
	\label{eq:nu}
\end{align}
The radiation component is~\citep{1962pfig.book.....S}
\begin{align}
	\nu_{\rm rad} = \frac{4 a T^4}{15 c \kappa \rho^2},
	\label{eq:nu_rad}
\end{align}	
where $a$ is the radiation gas constant.
We obtain the hydrogen viscosity using the Braginskii-Spitzer formula~\citep{osti_4317183,1962pfig.book.....S}
\begin{align}
	\nu_{\rm ie} \approx 2.21\times 10^{-15} \frac{(T/\mathrm{K})^{5/2}}{(\rho/\mathrm{g\,cm^{-3}}) \ln \Lambda}\mathrm{cm^2\,s^{-1}},
	\label{eq:nu_ie}
\end{align}
where $\ln \Lambda$ is the Coulomb logarithm, given for hydrogen by~\citep{1987ApJ...313..284W}
\begin{align}
	\ln \Lambda = -17.9 + 1.5\ln \frac{T}{\mathrm{K}} - 0.5 \ln\frac{\rho}{\mathrm{g\,cm^{-3}}}
\end{align}
for temperatures $T < 4.2\times 10^5\,\mathrm{K}$ and
\begin{align}
	\ln \Lambda = -11.5 + \ln \frac{T}{\mathrm{K}} - 0.5 \ln\frac{\rho}{\mathrm{g\,cm^{-3}}}
\end{align}
for $T > 4.2\times 10^5\,\mathrm{K}$.
Corrections owing to different compositions are generally small relative to the many orders of magnitude we are interested in here.
For instance the difference between pure hydrogen and a cosmic mixture of hydrogen and helium is under $30\%$~\citep{2008ApJ...674..408B}.

\section{Data Availability}

The plotting scripts used in this work, as well as the MESA run directories used in producing Figure~\ref{fig:prof}, may be found in this GitHub \href{https://github.com/adamjermyn/cbm_ra_plots}{repository} in the commit with short-sha d7cdf2f. The inlists and run scripts used in producing the HR diagrams in this work are available in this other GitHub \href{https://github.com/adamjermyn/conv_trends}{repository} on the main branch in the commit with short-sha 964e07d. The data those scripts produced are available in~\citet{jermyn_adam_2022_5878965}.

%% file: chandra.tex
\section{Critical Rayleigh Number}\label{appen:chandra}
The critical Rayleigh number depends on both the value of the Taylor number (rotation) and the Prandtl number (diffusivity ratio),
\begin{align}
	\mathrm{Ta} \equiv \frac{4 \delta r^4 \Omega^2}{\nu^2}
\end{align}
and
\begin{align}
	\mathrm{Pr} \equiv \frac{\nu}{\alpha}.
\end{align}
Recent work has explored the intricacies of convective onset in rotating, spherical domains \citep{2008JFM...601..317N, 2018PhRvF...3b4801G}.
For simplicity, we use the derivation of \cite{1961hhs..book.....C} to determine convective onset.
For a given value of Ta and Pr, we use his Ch.~3, eqns.~222 \& 227 to respectively calculate the critical Rayleigh number of an overstable onset and direct onset,
\begin{align}
&\mathrm{Ra}_{\mathrm{crit,over}} = 2 \pi^4 \frac{(1 + \mathrm{Pr})}{x} \left((1 + x)^3 + \frac{\mathrm{Pr}^2 \mathrm{Ta}/\pi^4 }{ (1 + \mathrm{Pr})^2 }\right), \\
&\mathrm{Ra}_{\mathrm{crit,direct}} = \pi^4\left( \frac{ (1 + x)^3 + \mathrm{Ta}/\pi^4 }{x}\right),
\end{align}
where $x$ is a nondimensional wavenumber of convective onset, which must be compared to a reference value (eqn.~225),
\begin{equation}
x^* = \left(\frac{\mathrm{Ta} (1 - \mathrm{Pr})}{\pi^4(1 + \mathrm{Pr})}\right)^{1/3} - 1.
\end{equation}
If $x < x^*$, $\mathrm{Ra}_{\mathrm{crit}} = \mathrm{Ra}_{\mathrm{crit,over}}$, otherwise $\mathrm{Ra}_{\mathrm{crit}} = \mathrm{Ra}_{\mathrm{crit,direct}}$.
For a given value of Ta and Pr, we evaluate $\mathrm{Ra}_{\mathrm{crit}}$ for $x \in [10^{-3}, 10]$, and take the minimum value to be the true value of $\mathrm{Ra}_{\mathrm{crit}}$.